\documentclass[a4paper,fleqn]{cas-sc}

\usepackage{geometry}
\RequirePackage[colorlinks]{hyperref}
\colorlet{scolor}{black}
\colorlet{hscolor}{DarkSlateGrey}

\usepackage[english]{babel}
\usepackage[round]{natbib}
\usepackage{url}
\usepackage{float}

\usepackage[shortlabels]{enumitem}

\def\tsc#1{\csdef{#1}{\textsc{\lowercase{#1}}\xspace}}
\tsc{WGM}
\tsc{QE}
\tsc{EP}
\tsc{PMS}
\tsc{BEC}
\tsc{DE}

\begin{document}
\let\WriteBookmarks\relax
\def\floatpagepagefraction{1}
\def\textpagefraction{.001}

\title [mode = title]{Housing Property Rights and Social Integration of Migrant Population: Based on the 2017 China Migrants' Dynamic Survey}

\tnotemark[1]
\tnotetext[1]{The authors thank the Migrant Population Service Center, National Health Commission P. R. China for data support.}

\author[1]{Jingwen Tan}[orcid=0000-0002-3452-959X]
\cormark[1]
\ead{tjw@henu.edu.cn}
\credit{Conceptualization of this study, Methodology, Software}
\affiliation[1]{organization={School of Economics, Henan University},
    addressline={Jinming Avenue}, 
    city={Kaifeng},
    postcode={475004}, 
    country={China}}
\author[1]{Shixi Kang}[orcid=0000-0003-4847-5171]
\cormark[2]
\ead{ksx@henu.edu.cn}
\credit{Conceptualization of this study, Methodology, Software}
\cortext[cor1]{Corresponding author}
\cortext[cor2]{Principal corresponding author}


\begin{abstract}
Push-pull theory, one of the most important macro theories in demography, argues that population migration is driven by a combination of push (repulsive) forces at the place of emigration and pull (attractive) forces at the place of emigration. Based on the push-pull theory, this paper shows another practical perspective of the theory by measuring the reverse push and pull forces from the perspective of housing property rights. We use OLS and sequential Probit models to analyze the impact of urban and rural property rights factors on the social integration of the migrant population based on “China Migrants' Dynamic Survey". We found that after controlling for personal and urban characteristics, there is a significant negative effect of rural property rights (homestead) ownership of the mobile population on their socio-economic integration, and cultural and psychological integration in the inflow area. The effect of urban house price on social integration of the migrant population is consistent with the "inverted U-shaped" nonlinear assumption: when the house price to income ratio of the migrant population in the inflow area increases beyond the inflection point, its social integration level decreases. That is,there is an inverse push force and pull force mechanism of housing property rights on population mobility.
\end{abstract}
\maketitle
\section{Introduction}
Urbanization is a necessary process when a developing country develops into a developed country \citep{zhang2016trends}. Since 1980s, China’s migrant size has increasingly expanded. Countryside’s outflow population provides important human resources to urban development and plays huge role in the urbanization process \citep{zhang2003rural}. This phenomenon arises mainly from the gradual saturation of the demand for human resources in the agricultural sector. It is estimated that since 2003, the agricultural sector can only support 1.7 million jobs, while the surplus labor force in rural areas reached 1.5 million \citep{Xinhua}. By 2020, the migrants in China were 375.82 million, increased by 69.73\% in the past 10 years \citep{NBS2021}.  In the policy environment of China's manufacturing development, the potential power of the mobile population for economic development is being increasingly appreciated.

However, in the urbanization process of China, migrants bear a low integration level on social insurance, cultural life and identity, and even a gap on the aspect of public resource allocation when comparing with the residents in urban destinations \citep{Yunsongchen2015Inequality}. In a study of the special treatment of migrants in the urban job market, \cite{meng2001two} found that there is a significant occupational discontinuity between urban and rural residents, and that this discontinuity cannot be explained by productivity differences. Urban residents enjoy more preferences and facilities in the job market than the migrants. Thus, migrants are really hard to integrate with the local society, certain scholars proposed the concept of “peri-urbanization” to exposit such phenomenan \citep{ChunguangWang2006Study}. As an important factor influencing the migration of mobile population to the local area, how to improve the willingness of this group to integrate has become an urgent task for cities to achieve human capital agglomeration.

The forces acting on the mobile population's willingness to integrate socially are mainly derived from the push and pull forces of the city. In the push-pull theory proposed by \cite{lee1966theory}, favorable factors in cities will attract people to move in and unfavorable factors will drive population loss. A view that is gaining ground is that the key factor preventing mobile populations from moving in is household registration, and \cite{solinger1999contesting} points out that the availability of legal and permanent household registration to immigrants is an important reason for their settlement decisions. Considering the reverse push-pull perspective, the advantage of the homestead will also be an important driving force to pull the population back. In an empirical study on the influence of the residence base on the migration decision of the migrant population in China, \cite{gu2020does} found that the property rights of migrants in their original residence can also be one of their resistance to move to the city. Since the existence of rural hukou is an important basis for allocating residential bases to migrants, once they move to the city, this group will inevitably suffer great losses because they can no longer own the property rights of their original residence. Therefore, the hindrance of moving into the household registration and the ownership of property rights in the place of outflow together constitute the push force that affects the willingness of social integration of the mobile population.

In this study, we followed the theory to explore the flow pattern of urban and rural migrants in China. By following the pull and push theory and adopting the dynamic monitoring data on China’s migrants in 2017, this paper explores the influence of the factors of urban and rural home ownership on migrants’ subjective social integration intentions.The rest of the paper is organized as follows: Section 2 is a literature review; Section 3 is the study design; Section 4 is the data and variables presentation; Section 5 is the empirical process and analysis; and Section 6 concludes and makes recommendations.
\section{Research background}

\begin{enumerate}[(1)]
\item Traditional measurement method of social integration indicator.
\end{enumerate}

The concept of social integration is the core of understanding the difference in the different groups in the contemporary society, especially the social community at the marginal status \citep{wang2012migrant}. Certain scholars proposed that the individual difference of migrants had impact on their social integration level. \cite{zhang2007assimilation} studied the gap in the income between migrants and urban populations based on the dynamistic perspective, and conducted the empirical study on the income assimilation of migrants in cities from the perspective of labor market. The study of \cite{wang2012migrant} showed that the migrant who was the self-employed individual capable of speaking local dialect and rich could integrate with local urban society better. Some studies focused on the influences of city characteristics and rural land on social integration. For instance, \cite{massey1990social} considered that the factor, namely the city characteristics of migrants’ urban destination were important to population migration, Besides, there are other factors such as city characteristics, rural land and policy etc.

Among the studies on the social integration of China’s migrants, \cite{tian2019impacts} also reckoned that the specific city factor could affect the status of urban destination. \cite{zheng2020affordable} studied and discovered that the migrants lived in the economically affordable housing showed higher social integration level than those migrants lived in other types of housing. \cite{tyner2016hukou} studied and discovered that the social integration level of migrants was related to the land ownership to some extent. \cite{zhang2020urban} discovered that the land-lost families migrated from countryside to cities did not integrate with local urban society better, but land acquisition improved their life quality and social welfare. Presently, most of studies focus on the influence of single city, town or countryside on the social integration level of migrant, but few on the measurement in both directions of countryside and cities.

Determining one uniform and effective social integration measurement dimension and indicator system is playing vital role in the process of studying migrants’ integration \citep{entzinger2003benchmarking}. According to the classical theory of social integration measurement dimension, \cite{park1928human} defined the social integration as the interpenetration and assimilation of single-dimension individual or groups between cultures. \cite{gordon1964assimilation} established the “two dimensions” model of social integration theory against two aspects, namely cultural fusion and structural assimilation. \cite{entzinger2003benchmarking} based on the subjective and objective social integration intentions of individual, put forward to a rather comprehensive measurement dimension from various aspects such as society, economy, culture, law and politics. \cite{basu2013group} elaborated innovatively the importance of identity for social integration on its basis.

All literatures aforesaid considered the cultural fusion as the important integral part of social integration. Cultural fusion is always a key consideration whichever a dimension is adopted to measure social integration degree. However such thought faces with certain limitations in the concrete practice. For example, the concept of cultural fusion is not clear. Some studies brought the sense of belonging and identity of migrants to urban destinations into the scope of cultural fusion, and some mingled them with psychological fusion, so the boundaries of cultural fusion and psychological fusion are rather fuzzy. For another example, the progressive relationship of indicators such as economic amalgamation, social integration and cultural fusion is stressed, but the possible interactions among various factors are neglected. \cite{basu2013group} Reckoned that identity could effectively improve the performance of laborer on the labor market and further promote economic amalgamation. However economic amalgamation may also generate the counteraction to the group identification sense of migrants.

According to the studies aforesaid, economy, culture, politics and identity constitute the major indicator system of social integration. Currently, most of studies on the influencing factors of social integration willingness focus on the analysis on social capital, social discrimination, human capital and labor market \citep{Ren2006Social}. As the theories related to social integration willingness is kept increasing and the statistical data are diversifying, there are huge spaces to develop the traditional measurement standards and indicator systems.

This paper constitutes the following measurement indicator systems on the premise of clarifying every indicator concept:

Socio-economic amalgamation indicator: 

Is the migrant willing to settle in the local area when meeting the local settlement conditions (econ\_1)? 

Is the migrant willing to reside locally to work and live for a while (econ\_2)?

Cultural and psychological fusion indicator:

Does the migrant treat himself/herself as a native (psyc\_1)? 

Is the migrant more willing to work or handle affairs by referring to local living habits (psyc\_2)?

\begin{enumerate}[(2)]
\item Analysis on subjective social integration willingness based on the pull and push theory.
\end{enumerate}

The pull and push theory is the important theory in the category of population migration.\cite{lee1966theory} as proposed, the factor benefiting population inflow in the outflow area is the pulling force, but the factor unfavoring to the outflow area is the pushing force. According to the neoclassical economics, the possibility of migration is arose from the one that migration benefit is higher than migration cost \citep{bogue}. According to existing literatures, city’s pulling force and countryside's pushing force will have positive impact on the subjective social integration willingness of migrants.\cite{mcgranahan2002understanding} studied and found that the incidence of poverty in much countryside in USA started increasing and populations reduced almost one fourth due to single industrial structure and imperfect infrastructure construction. \cite{cromartie2001nonmetro} upon the current CPS (Current Population Survey), found that 1.90 million populations migrated to big cities from the non-city areas in one year till March 2000. According to the theory of territorial divisions of labour \citep{lipietz1977capital,massey1995spatial,scott1988metropolis}, city is the place for assembling various industries in space, it provides another integration to the one that city pulls population inflow and improves their integration willingness.

However city’s counter pushing force and countryside's counter pulling force measure the subjective social integration willingness of migrants from a reverse angle. Whilst migrants move to cities in a large scale, some labor forces may flow back on the “compensation” basis \citep{van1985discrepancy}, and \cite{labrianidis2009geographical}surveyed and found that the migrants in developed countries were as high as 50\%. For this phenomena, \cite{dustmann1996return} gave explanations from the aspects of comparative gain, accumulative ROI (return on investment) and personal preference. 

On one hand, the counter pushing force of city blocks the population inflow. \cite{berry2008urbanization} found in the study of urbanization process, one major reason of urban population outflow is the higher living cost. \cite{wang2020driving} thought that the worsening urban employment environment resulted from single industrial structure was the important reason why migrant population returned back. \cite{chan2010household} studied and found that China’s household registration system under the dualistic system limited migrant population to enjoy the rights of public services and taking part in politics, it explained the reason why there was urban population outflow from the level of institutional system.

On the other hand, the counter pulling force in countryside reduces the willingness of migrant population for integrating with cities. \cite{davanzo1983repeat} by introducing the concept of “location specific capital”, proposed the viewpoint that the more human capital accumulated by migrants in a certain area is, the larger possibility that migrants return to the area is. Undoubtedly, the existence of rural housing and land, being the important representatives of capital accumulation, weakens the willingness of migrant population for actively integrating with cities \cite{hu2011circular}. There is another explanation derived from the Target Income Theory \cite{piore1979birds}, namely the improvement of economic condition can shorten the dwell time of migrant population in the urban destinations. Its reason is the one that the migration between countryside and city is essentially a family strategy accelerating capital accumulation, its purpose is to create conditions for acquiring better development opportunities after migrants return back \citep{vadean2009circular,constant2002return}. In the retirement model of \cite{henderson2005aspects}, people are usually inclined to return to countryside after completing necessary capital accumulation in cities, and the phenomena that migrants prefer to consume in their original places \citep{hill1987immigrant,djajic1988general} become the rural counter pulling force of migrants for returning back.

Presently, most of studies on the push and pull theory in the academic circle are concentrated on the mobility of rural labor forces, and few of them focus on the analysis of counter pulling and pushing force. In this paper, we focus on measuring the impact of counter pulling and pushing force on the social integration of migrant population.

\section{Research design}
\begin{enumerate}[(1)]
\item Counter pulling force measurement of rural ownership right on migrant population.
\end{enumerate}

China has a unique population management system. The household registration system(hukou) is a household-based population management system implemented for its own citizens settled in mainland China, which divides household attributes into agricultural and non-agricultural households based on geographic and family member relationships, indicating the legitimacy of natural persons living locally. This dualistic urban-rural household registration system inhibits the free movement of urban and rural populations and is discriminatory in terms of social welfare.

Homestead are land occupied by Chinese rural families as residential bases, and the property rights belong to rural collectives. Rural hukou holders lose their inheritance rights to the homestead if they move their hukou to the city, i.e., the residential land becomes an asset of the village collective.  In this paper, the research hypothesis considers that the ownership of residential land of mobile population affects their willingness to integrate locally.

Contracted land refers to the right of members of rural collective economic organizations to contract rural land issued by the collective economic organizations according to law, and the contractors enjoy the right to use, gain and transfer the contracted land management rights according to law, which is a kind of economic rights and interests. The nature of property rights of contracted land is similar to that of residential bases, which is owned by rural collectives. In this paper, we use contracted land as a proxy variable for homestead base in the robustness test section for regression.

In this paper, the explained variable, i.e.: the social integration indicator of migrant population is the non-continuity variable, the availably optional model includes ordered Probit model, multi-class Logistic model etc. In the linear fitting process, the Least Square Method, compared with the estimation methods such as likelihood estimation etc., has less hypothesis, so the estimated parameter fitting results on the statistical sense is more robust. According to the study of \cite{nunn2011slave}, when the explained variable is a dummy variable or an ordinal variable, the OLS regression method can still be used. In this paper, we adopted the Linear Probability Model (LPM) as the major regression model, and the ordered Probit model to conduct the robustness test. The basic expression of LPM model is:
\newtheorem{theorem}{Theorem}
\begin{theorem}
\begin{eqnarray}\label{1}
S_{i}^{k}=\alpha+\beta \ln L_{i}+X_{i}^{\prime} \gamma+\varepsilon_{i}, \varepsilon_{i} \sim N\left(0, \quad \sigma^{2}\right)
\end{eqnarray}
\end{theorem}
Where: $S_{i}^{k}$ is the Kth type of subjective social integration willingness of the sample i, $L_{i}$ is the area or contracted land and house site owned by rural resident i, $X_{i}$ is the control variables.

The basic expression of Probit model is:
\begin{theorem}
\begin{eqnarray}\label{2}
\operatorname{Pr}\left(\mathrm{S}_{i j}=1 \mid h i, X\right)=F\left(h i, \beta_{1}\right)=\frac{\exp \left(\beta_{0}+\beta_{1} h i_{i j}+\beta X+\varepsilon_{i j, t}\right)}{1+\exp \left(\beta_{0}+\beta_{1} h i_{i j}+\beta X+\varepsilon_{i j, t}\right)}
\end{eqnarray}
\end{theorem}
Where: ${S}_{i j}$ is the intention of social integration of the mobile population, a binary variable. ${Pr}({S}_{i j}=1h i, X)$ is the probability that the mobile population has the intention of social integration in the inflow city. F is the cumulative distribution function of the standard normal. $h_{i}$ is the home base indicator, X is other control variables, and $\varepsilon_{i j, t}$ is a random disturbance term.

\begin{enumerate}[(2)]
\item Counter pulling force measurement of urban ownership right on migrant population.
\end{enumerate}
The willingness of the mobile population to integrate in the inflow area also depends on the difficulty of acquiring urban housing property rights. Even if the migrant population is not affected by the ownership of rural residential property rights, high urban housing prices are an obstacle to their long-term living in the inflow area. In this section, we studied the housing price / the disposable personal income 2017, i.e.: the core explanatory variable. The basic expression of LPM model is:

\begin{theorem}
\begin{eqnarray}\label{3}
S_{i}^{k}=\alpha+\beta H_{t}+X_{i}^{\prime} \gamma+\varepsilon_{i}, \varepsilon_{i} \sim N\left(0, \quad \sigma^{2}\right)
\end{eqnarray}
\end{theorem}

Where: S is the kth type of subjective social integration willingness of the sample i, Ht is the ratio of housing price and income in the sample city, X’i is the control variable. Simultaneously, the quadratic-term estimate result of housing price ~ income ratio was added in this section, we thereby analyzed the break point of the impact of the change in housing price ~ income ratio on the subjective integration willingness of migrant population.

\begin{enumerate}[(3)]
\item Bias error of sample selection: Propensity Score Matching.
\end{enumerate}

Despite the wide coverage and large sample size of the CMDS database used in this paper, and the inclusion of individual control variables in the model, there is still a certain problem of sample selection bias in the ownership of homesteads and contracted land by the mobile population - the mobile behavior from the rural population may itself be related to the ownership of production and means of livelihood, i.e., the endogenous problem of "self-selection". The Propensity Score Matching is the preferred method for the robustness test, it can effectively relieve the problem.

In this paper, we adopted the 1:2 nearest neighbor matching method to analyze, the radius matching and the kernel matching to conduct the robustness test.
\begin{enumerate}[(4)]
\item Bidirectional causality: Instrumental variable.
\end{enumerate}

The core explanatory variable of the measurement on the impact of urban ownership on migrant’s subjective integration willingness is the housing price ~ income ratio. The housing price is a highly endogenous variable and is inclined to be affected by multiple factors. When the social integration degree measurement indicator of migrant in a certain area is higher, the potential demand may possibly push up housing price, so the bidirectional causality endogenesis problem may generate in the predication and which may affect the unbiasedness of estimation. In the process of treating bidirectional causality problem, the instrumental variable is a widely applied method.

Common instrumental variables in empirical studies targeting house prices include land development area, the product of long-term interest rate and land supply elasticity, and land sales price \citep{Dongdongpeng2016Do,WeiZhang2018HousePrice,chaney2012collateral,waxman2020tightening}. In this paper, we choose lagged land price as the instrumental variable of house price. Land sales price directly affects house prices, which is consistent with the hypothesis of correlation of instrumental variables; meanwhile, mobile population does not consider land sales price in the process of integration in the inflow, which is consistent with the hypothesis of exogeneity of instrumental variables.

In our empirical study, 5\% Wald Test on the nominal significance level of instrumental variable, in principle, can refuse the hypothesis of weak instrumental variable. In order to guarantee the reliability of conclusions, we simultaneously selected the maximum likelihood method of limited information that is more accurate for predicating weak instrumental variable to make regression. Besides, if the disturbing term has heteroscedasticity or self-correlation, GMM (Generalized method of moments) is more effective relative to 2SLS. In order to prevent the potential disturbing term from affecting the estimated unbiasedness, GMM was added to conduct the robustness test.

It is worth adding that we do not believe that there is a causal problem with the mechanism of action of the first part of the study(homestead on the social integration of the mobile population). The mobile population's willingness to integrate in the inflowing place does not affect all the cases of their home bases and contracted land. Only when mobile people give up their rural hukou and transfer to urban hukou, their home base ownership changes, and the sample selected for this paper is a mobile group that has not acquired a household registration in the inflowing place.

\section{Date introduction and descriptive statistics}
\begin{enumerate}[(1)]
\item Data introduction.
\end{enumerate}

In our study, the dynamic monitoring survey data of the National Health Commission of the People’s Republic of China on migrant populations 2017 were selected to conduct the empirical analysis. The survey covered 31 provinces (municipalities, autonomous regions) of China, where those migrants who stayed in the local area for one or more months were selected to conduct the dynamic monitoring and survey. The average age of samples under survey was 15~59 years old, it adopted the hierarchical, multi-stage and large-scale PPS sampling method to conduct a careful survey on Chinese migrant’s development, individual characteristic, social integration and employment, the results therefrom provide data support to the study of social science in the directions of labor economics, demography, etc.
\begin{enumerate}[(2)]
\item Variable and descriptive statistics.
\end{enumerate}

\begin{table}[htb]
\caption{Baseline regression results. }\label{tbl1}
\begin{tabular*}{\tblwidth}{@{} CCCCCCC@{} }
\toprule
variables &deperssion                                         & N       & mean    & sd      & min    & max       \\
\midrule                        
lnzjd     &logarithmic homestead area                            & 153,272 & 2.241   & 2.014   & 0      & 6.908     \\
lncbd     &logarithmic contracted land area                      & 153,272 & 2.993   & 3.379   & 0      & 13.05     \\
hosin     &house price to income ratio                           & 153,115 & 0.249   & 0.127   & 0.0861 & 0.906     \\
\midrule                       
econ\_1   &willingness to obtain hukou                           & 153,272 & 0.394   & 0.489   & 0      & 1         \\
econ\_2   &willingness to stay and live in the area              & 153,272 & 0.831   & 0.374   & 0      & 1         \\
pscy\_1   &consider yourself a local                             & 153,272 & 0.751   & 0.433   & 0      & 1         \\
pscy\_2   &follow the local habits                               & 153,272 & 0.444   & 0.497   & 0      & 1         \\
\midrule                      
teacher   &number of teachers per 1,000 people                   & 146,632 & 19,960  & 20,685  & 672    & 86,957    \\
book      &book collection per capita                            & 146,712 & 103.4   & 173.0   & 0.100  & 777.3     \\
doctor    &number of physicians per 1,000 people                 & 141,473 & 20,792  & 23,289  & 469    & 94,417    \\
pgdp      &GDP per capita                                        & 141,073 & 101,701 & 154,625 & 17,890 & 6.422e+06 \\
tgdp      &proportion of tertiary industry employees             & 145,633 & 57.17   & 11.20   & 27.71  & 80.56     \\
gende     &gender                                                & 153,272 & 0.514   & 0.500   & 0      & 1         \\
scope     &scope of mobility                                     & 153,272 & 2.305   & 0.756   & 1      & 3         \\
edu       &education years                                       & 153,272 & 10.21   & 3.386   & 0      & 19        \\
married   &marital status                                        & 153,272 & 0.813   & 0.390   & 0      & 1         \\      
\bottomrule
\end{tabular*}
\end{table}

For the explained variable concerned by the paper, namely the social integration indicator, we followed the indicator system proposed by \cite{entzinger2003benchmarking}and \cite{basu2013group} and selected the economic amalgamation, cultural and psychological fusion as the measurement basis. In the CMDS questionnaire, the corresponding questions are: “if you meet the local settlement conditions, are you willing to change your residence registration to the local area?” “Are you planning to continue to stay at the local area in your future time?” “Do you agree the saying that ‘I am feeling that I am a native person’” “Do you agree ‘it is important to me if handling affairs by following the manners and customs of original place?’”.

In this paper, we select the home base ownership of the mobile population and the house price to income ratio of the inflowing city as the proxy variables of rural "counter-push" and urban "counter-pull".

The demographic variable is usually used as the control variables in the empirical study. In the human capital theory, these factors, such as educational level, work experience etc. can help individuals to acquire better jobs, higher incomes, so they are capable of integrating with developed regions. In the relational demography theory, certain characteristic homogeny and difference of individual and group may affect the satisfaction of interpersonal interaction and the related attitude and behavior. In this paper, we referred to the theories aforesaid and selected four indicators, namely gender, education years, scope of mobility and marital status, as the control variables.

City characteristics may affect the level of social integration of the migrant population in the inflow area. In order to ensure the accuracy of estimation, a reasonable selection of city control variables is required. The city control variables selected in this paper mainly reflect economic, industrial structure and infrastructure characteristics. They mainly include the proportion of tertiary industry employees, GDP per capita, book collection per capita, number of physicians per 1,000 people, and number of elementary school teachers per 1,000 people.

\section{Empirical process and discussion}
\begin{enumerate}[(1)]
\item Impact of rural ownership on migrant’s integration willingness.
\end{enumerate}

\begin{table}[htb]
\caption{Baseline regression results. }\label{tbl2}
\begin{tabular*}{\tblwidth}{@{} CCCCC@{} }
\toprule
            & (1)           & (2)            & (3)            & (4)             \\
            & econ\_1       & econ\_2        & psyc\_1        & psyc\_2         \\
\midrule            
lnzjd       & -0.0243***    & -0.00156***    & -0.0152***     & -0.0162***      \\
            & (0.000650)    & (0.000515)     & (0.000582)     & (0.000685)      \\
married     & 0.0347***     & 0.0896***      & 0.0719***      & -0.0369***      \\
            & (0.00338)     & (0.00292)      & (0.00317)      & (0.00355)       \\
sex         & -0.0102***    & -0.00654***    & 0.00698***     & -0.0275***      \\
            & (0.00251)     & (0.00201)      & (0.00232)      & (0.00267)       \\
edu         & 0.0156***     & 0.0108***      & 0.00477***     & 0.0182***       \\
            & (0.000396)    & (0.000311)     & (0.000357)     & (0.000415)      \\
scope       & 0.00227       & -0.0223***     & -0.0877***     & -0.00142        \\
            & (0.00175)     & (0.00147)      & (0.00155)      & (0.00189)       \\
teacher     & 0.00000333*** & 0.000000187    & -0.00000497*** & -0.000000809*** \\
            & (0.000000246) & (0.000000184)  & (0.000000242)  & (0.000000252)   \\
book        & 0.000114***   & 0.0000738***   & 0.0000593***   & 0.0000513***    \\
            & (0.0000116)   & (0.00000828)   & (0.0000122)    & (0.0000127)     \\
doctor      & 0.00000205*** & 0.000000557*** & 0.00000154***  & 0.000000987***  \\
            & (0.000000159) & (0.000000119)  & (0.000000166)  & (0.000000172)   \\
tgdp        & 0.00219***    & 0.000142       & 0.00110***     & -0.000302*      \\
            & (0.000148)    & (0.000124)     & (0.000128)     & (0.000158)      \\
pgdp        & -6.52e-09     & 2.17e-08***    & -1.48e-08**    & 2.04e-08**      \\
            & (7.24e-09)    & (5.92e-09)     & (6.49e-09)     & (8.24e-09)      \\
cons        & 0.0224**      & 0.672***       & 0.856***       & 0.340***        \\
            & (0.0103)      & (0.00854)      & (0.00897)      & (0.0110)        \\
\midrule            
N           & 137033        & 137033         & 137033         & 137033          \\
$R^{2}$     & 0.115         & 0.024          & 0.052          & 0.028           \\
\bottomrule
\end{tabular*}
\end{table}

In the first part of empirical study, the impact of migrant’s rural property rights ownership on his/her subjective willingness of integrating with local urban destination. According to the regression result as shown in Table 2, the homestead owned by migrant may generate significant impact on his/her subjective willingness of integrating with local urban destination.Specifically, the willingness to obtain a household registration in the inflow area is 2.43\% lower for the mobile population with a homestead than for those without a homestead, and the willingness to stay is 0.156\% lower. When the mobile population has a homestead, the probability of considering themselves as residents of the inflow area will be 1.52\% lower and the probability of following the living habits of the inflow area will be 1.62\% lower than that of the group without a homestead.

The apparent negative correlation between mobile population ownership of the home base and the four indicators in the benchmark regression reflects an important economic issue. In the discussion of the relationship between home base ownership and social and economic integration indicators, the argument relies on the simple logic that fixed assets such as property are important for the development of people's lives, so that home purchase in the inflowing area becomes a key characteristic of the mobile population in deciding to stay*. When the mobile population owns a home in another region, the property purchased in the inflow location loses its most important residential attribute. The majority of the mobile population's willingness to purchase housing originates from residence rather than investment*, so once this happens, the possibility of long-term residence for the mobile population decreases as the willingness to purchase housing locally decreases. Meanwhile, in China, hukou is seen as a kind of threshold for home ownership. Based on the previous logic, the mobility population's willingness to settle down will also decrease in tandem with the willingness to purchase a home.

Moreover, this paper then explores the negative relationship between home base ownership and cultural and psychological integration indicators. This relationship still takes the willingness to purchase a home locally as an intermediate medium of action. According to the results discussed in the previous paragraph, ownership of the home base directly leads to a decrease in the willingness of the mobile population to purchase a home in the local area, which means that this segment of the mobile population prefers short-term housing of a rental nature. The emergence of this tendency will make it easier for the mobile population to congregate in similar groups as opposed to entering the residential areas of local residents. As a result, they will also have difficulty in gaining cultural and psychological identity.

\begin{table}[htb]
\caption{Fixed effects regression. }\label{tbl3}
\begin{tabular*}{\tblwidth}{@{} CCCCC@{} }
\toprule
                  & (1)        & (2)        & (3)         & (4)        \\
                  & econ\_1    & econ\_2    & pscy\_1     & pscy\_2    \\
\midrule                   
lnzjd             & -0.0226*** & -0.000864  & -0.00922*** & -0.0125*** \\
                  & (0.000665) & (0.000533) & (0.000601)  & (0.000708) \\
Control variables & $\surd$    & $\surd$    & $\surd$     & $\surd$    \\
Fixed effect      & $\surd$    & $\surd$    & $\surd$     & $\surd$    \\
\midrule 
N                 & 137033     & 137033     & 137033      & 137033     \\
$R^{2}$           & 0.150      & 0.045      & 0.093       & 0.055      \\
\bottomrule
\end{tabular*}
\end{table}

\begin{table}[htb]
\caption{Probit model regression. }\label{tbl4}
\begin{tabular*}{\tblwidth}{@{} CCCCC@{} }
\toprule
                  & (1)             & (2)             & (3)             & (4)             \\
                  & Probit\_econ\_1 & Probit\_econ\_2 & Probit\_pscy\_1 & Probit\_pscy\_2 \\
\midrule                   
lnzjd             & -0.0683***      & -0.00811***     & -0.0502***      & -0.0417***      \\
                  & (0.00183)       & (0.00210)       & (0.00192)       & (0.00177)       \\
Control variables & $\surd$         & $\surd$         & $\surd$         & $\surd$         \\
Fixed effect      & -               & -               & -               & -               \\
\midrule 
N                 & 137033          & 137033          & 137033          & 137033          \\  
\bottomrule
\end{tabular*}
\end{table}

\begin{table}[htb]
\caption{Contracted land regression results. }\label{tbl5}
\begin{tabular*}{\tblwidth}{@{} CCCCC@{} }
\toprule
                  & (1)          & (2)          & (3)          & (4)          \\
                  & OLS\_econ\_1 & OLS\_econ\_2 & OLS\_pscy\_1 & OLS\_pscy\_2 \\
\midrule                   
lncbd             & -0.0116***   & 0.00150***   & -0.00444***  & -0.00383***  \\
                  & (0.000389)   & (0.000313)   & (0.000356)   & (0.000413)   \\
Control variables & $\surd$      & $\surd$      & $\surd$      & $\surd$      \\
Fixed effect      & $\surd$      & $\surd$      & $\surd$      & $\surd$      \\
\midrule 
N                 & 137033       & 137033       & 137033       & 137033       \\
$R^{2}$           & 0.148        & 0.046        & 0.092        & 0.053        \\
\bottomrule
\end{tabular*}
\end{table}

To make the estimation results more precise, this paper performs robustness tests by adding urban fixed effects (Table 3), Probit model (Table 4) and replacing core explanatory variables (Table 5), respectively. Controlling for urban fixed effects through dummy variables is prone to the problem of multicollinearity, which affects the significance of the results. Table 3 regression results controlling for urban fixed effects show insignificant socioeconomic integration indicators. The regression results of the probit model in Table IV are generally consistent with the economic significance of the OLS results. Table V shows changes in the coefficients of indicators of willingness to stay and work and live locally after replacing the core explanatory variable with contracted land. The economic significance of contracted land is more pronounced relative to the housing property rights represented by the homestead. The mobile population group leaving the countryside is itself insensitive to the agricultural production relationship represented by the homestead. Also, the inclusion of fixed effects has a shock to the regression coefficients. The results of the three robustness tests yield essentially the same conclusions as the baseline regression results.

\begin{enumerate}[(2)]
\item Robustness tests: propensity score matching.
\end{enumerate}

Propensity score matching can effectively mitigate the endogeneity problem caused by sample selection bias. In this paper, we use the matched sample for analysis to estimate the effect of matched mobile population Homestead ownership on their willingness to integrate in the city after common support hypothesis testing and stationarity testing. Table 6 shows the PSM estimation results, and the matching results are basically consistent with the OLS results while controlling for the individual characteristics of the sample. The homestead property ownership status of the mobile population in villages still has a significant negative effect on the socio-economic integration and cultural-psychological integration of the mobile population.
\begin{table}[htb]
\caption{Propensity score matching. }\label{tbl6}
\begin{tabular*}{\tblwidth}{@{} CCCCCCCC@{} }
\toprule
        & \multirow{2}{*}{Matching Status} & \multicolumn{2}{l}{1$\colon$1 Matching} & \multicolumn{2}{l}{Radius Matching} & \multicolumn{2}{l}{Kernel Matching} \\
        &                                  & ATT              & t            & ATT                & t              & ATT                & t              \\
\midrule         
econ\_1 & Before matching                  & -0.139***        & -53.03       & -0.139***          & -53.03         & -0.139***          & -53.03         \\
        & After matching                   & -0.073***        & -21.96       & -0.072***          & -21.34         & -0.073***          & -24.67         \\
\midrule        
econ\_2 & Before matching                  & -0.008***        & -4.02        & -0.008***          & -4.02          & -0.008***          & -4.02          \\
        & After matching                   & 0.008***         & 3.29         & 0.007***           & 4.43           & 0.008***           & 3.78           \\
\midrule        
pscy\_1 & Before matching                  & -0.031***        & -13.16       & -0.031***          & -13.16         & -0.031***          & -13.16         \\
        & After matching                   & -0.022***        & -7.38        & -0.021***          & -8.46          & -0.021***          & 8.33           \\
\midrule        
pscy\_2 & Before matching                  & -0.070***        & -26.16       & -0.070***          & -26.16         & -0.070***          & -26.16         \\
        & After matching                   & -0.020***        & -6.08        & -0.028***          & -6.45          & -0.026***          & -7.03          \\
\bottomrule        
\end{tabular*}
\end{table}

\begin{enumerate}[(3)]
\item The influence of urban property rights factors on the subjective social integration intentions of the mobile population.
\end{enumerate}

Table 7 reports the estimated results of the house price to income ratio on the mobile population's willingness to integrate socially, and the overall effect is negative, i.e., the higher the house price to income ratio in a given place, the lower the mobile population's willingness to integrate socially in that area. However, it is obvious that only PSCY1 has a confidence level of 10\%. As a vital lifetime investment, property plays an indispensable role for residents to carry out various development and economic activities in the future. A higher house price to income ratio means that housing This possibility will have two consequences: on the one hand, residents will be unable to integrate economically into the local area due to the higher cost of living; on the other hand, the high housing cost will also force the mobile population to choose cheaper residential areas, thus increasing the aggregation among the flow population, reducing their opportunities for contact with locals, and will further weaken their identification with the culture of the place they move in.

\begin{table}
\caption{House price regression results. }\label{tbl7}
\begin{tabular*}{\tblwidth}{@{} CCCCC@{} }
\toprule
                   & (1)          & (2)          & (3)          & (4)          \\
                   & OLS\_econ\_1 & OLS\_econ\_2 & OLS\_pscy\_1 & OLS\_pscy\_2 \\
hosin              & -0.000737    & 0.0000270    & -0.000831**  & -0.000116    \\
                   & (0.000568)   & (0.000195)   & (0.000359)   & (0.000309)   \\
Control variables  & $\surd$      & $\surd$      & $\surd$      & $\surd$      \\
\midrule             
N                  & 136708       & 136708       & 136708       & 136708       \\
$R^{2}$            & 0.022        & 0.107        & 0.028        & 0.025        \\
\bottomrule
\end{tabular*}
\end{table}

\begin{enumerate}[(4)]
\item Quadratic term estimates for urban property rights.
\end{enumerate}

The impact of house price to income ratio on social integration of the mobile population was estimated above using a linear probability model, but is the impact of house price to income ratio on the measures always linear? Table 8 report the nonlinear relationship between the house price to income ratio and social integration of the mobile population. According to Table 8 we can calculate the inflection points of the house price to income ratio in the four models at 746, 561, 857, and 627, respectively. As can be seen, the primary and secondary coefficients show diametrically opposed mean negative and mean positive characteristics, respectively. In the short run, an increase in the house price to income ratio raises the survival cost of migrants and thus reduces their willingness to move in, while in the long run perspective, a high house price to income ratio implies a possible increase in expected income, which leads to a subjective perception of the mobile population that they can integrate locally at the social and economic levels. At the same time, as an important symbol of the city's economic level, a higher house price to income ratio reflects to some extent the prosperity of the city's economy, which makes the migrant population more inclined to identify psychologically with the city's culture.

Table 9 depicts the regression results after adding instrumental variables to the model. Upon comparison, it can be seen that the results of 2SLS are similar to the OLS regression results with a change in the magnitude of the coefficients, indicating that the potential endogeneity problem tends to underestimate the impact of house prices on social integration of the mobile population. In addition, the regression results of LIML and GMM are consistent with 2SLS, demonstrating that the possible heteroskedasticity interference problems of weak instrumental variables and potential nuisance terms in the empirical evidence have no significant effect on the estimation results.

\begin{table}
\caption{Quadratic term estimate. }\label{tbl8}
\begin{tabular*}{\tblwidth}{@{} CCCCC@{} }
\toprule
                  & (1)           & (2)           & (3)           & (4)           \\
                  & OLS\_econ\_1  & OLS\_econ\_2  & OLS\_pscy\_1  & OLS\_pscy\_2  \\
hosin2            & 0.00000231*** & 0.000000553*  & 0.00000154*** & 0.00000129*** \\
                  & (0.000000445) & (0.000000332) & (0.000000490) & (0.000000491) \\
hosin             & -0.00345***   & -0.000621     & -0.00264***   & -0.00162**    \\
                  & (0.000706)    & (0.000520)    & (0.000776)    & (0.000792)    \\
Control variables & $\surd$      & $\surd$      & $\surd$      & $\surd$          \\    
\midrule                    
N                 & 136708        & 136708        & 136708        & 136708        \\
$R^{2}$           & 0.022         & 0.107         & 0.028         & 0.025         \\
\bottomrule
\end{tabular*}
\end{table}

\begin{table}[htb]
\caption{IV Quadratic term estimate. }\label{tbl9}
\begin{tabular*}{\tblwidth}{@{} CCCCCC@{} }
\toprule
          &             & (1)           & (2)           & (3)           & (4)           \\
          &             & OLS           & TSLS          & LIML          & GMM           \\
\midrule              
Econ1     & hosin2      & 0.00000231*** & 0.00000294*** & 0.00000294*** & 0.00000294*** \\
          &             & (0.000000445) & (0.000000670) & (0.000000670) & (0.000000670) \\
          & hosin       & -0.00345***   & -0.00418***   & -0.00418***   & -0.00418***   \\
          &             & (0.000706)    & (0.000799)    & (0.000799)    & (0.000799)    \\
\textit{} & $R^{2}$     & 0.022         & 0.022         & 0.022         & 0.022         \\
\midrule    
Econ2     & hosin2      & 0.000000553*  & 0.000000714   & 0.000000714   & 0.000000714   \\
          &             & (0.000000332) & (0.000000436) & (0.000000436) & (0.000000436) \\
          & hosin       & -0.000621     & -0.000809     & -0.000809     & -0.000809     \\
          &             & (0.000520)    & (0.000611)    & (0.000611)    & (0.000611)    \\
\textit{} & $R^{2}$     & 0.107         & 0.107         & 0.107         & 0.107         \\
\midrule    
Pscy1     & hosin2      & 0.00000154*** & 0.00000236*** & 0.00000236*** & 0.00000236*** \\
          &             & (0.000000490) & (0.000000771) & (0.000000771) & (0.000000771) \\
          & hosin       & -0.00264***   & -0.00360***   & -0.00360***   & -0.00360***   \\
          &             & (0.000776)    & (0.000909)    & (0.000909)    & (0.000909)    \\
\textit{} & $R^{2}$     & 0.028         & 0.028         & 0.028         & 0.028         \\
\midrule    
Pscy2     & hosin2      & 0.00000129*** & 0.00000125*   & 0.00000125*   & 0.00000125*   \\
          &             & (0.000000491) & (0.000000672) & (0.000000672) & (0.000000672) \\
          & hosin       & -0.00162**    & -0.00158      & -0.00158      & -0.00158      \\
          &             & (0.000792)    & (0.00101)     & (0.00101)     & (0.00101)     \\
\textit{} & $R^{2}$     & 0.025         & 0.025         & 0.025         & 0.025         \\
\midrule   
\textit{} & N           & 136708        & 136708        & 136708        & 136708        \\
\bottomrule
\end{tabular*}
\end{table}

\section{Conclusion}

The size of China's mobile population continues to expand, and its social integration has attracted widespread attention. Based on the dynamic monitoring data of China's mobile population, this study uses OLS and sequential Probit models to analyze the influence of urban and rural property rights factors on the subjective integration intention of the mobile population. Through propensity score matching, this paper conducts robust type tests on the regression results of rural residential property rights affecting the subjective integration intention of the mobile population; we calculate the inflection point of house price to income ratio on the integration intention of the mobile population through quadratic term regression, while using instrumental variables to mitigate the effect of potential endogeneity on the regression results. The results show that there is a significant negative effect of rural residential property rights (home base) ownership on the socio-economic integration and cultural-psychological integration of the mobile population in the inflow area. The effect of urban house price on social integration of the mobile population is consistent with the nonlinear assumption - Within reality, there is a negative correlation between the willingness to socially integrate the mobile population in housing prices.

Based on the above results, this paper discusses in three directions.

\begin{itemize} 
\item Sound housing policies for the mobile population and increased public capital investment.
\end{itemize}

Enhancing public capital investment in housing and expanding individuals' access to housing resources will have a positive impact on the economic development and psychological satisfaction of the mobile population. Excessive housing prices will increase the cost of residence for the migrant population and reduce their subjective economic status, which in turn will have a negative impact on cultural and psychological integration. The government should invest more public capital in community construction, such as public housing for the migrant population, to foster community culture and thus promote the cultural and psychological integration of the migrant population.

\begin{itemize}

\item While implementing the reform of the household registration system, the government should take into account the real needs of the "two-way mobility" group.

\end{itemize}

As there are two-way flows in the process of urban-rural integration, individual differences among the floating population are obvious. On the one hand, we should actively solve the issue of household registration for the floating population and provide adequate protection for basic public services such as medical care, economy, and children's education; on the other hand, we should also face the contradiction between moving household registration and insufficient economic strength of some floating population, avoid secondary movements due to low income and unstable employment after transferring to the household registration, create a stable source of manpower for economic development, and promote sustainable urban economic and social development.

\begin{itemize}\item Strengthen urban management construction, coordinate industrial structure and create jobs.
\end{itemize}

To a certain extent, the level of urban management reflects the good or bad industrial planning and the degree of economic development. Strengthening urban management construction will effectively improve the efficiency of urban resource allocation and operation, and gain positive perception of mobile population. In order to improve the level of urban management, it is necessary to build an employment service system for the floating population, establish an effective training and management system for the floating population, and effectively promote the minimum wage system, thus gathering human capital for economic development and stimulating new dynamic energy for economic development.
\newpage

\bibliography{cas-refs}

\end{document}